\documentclass[]{pasj01}
\onecolumn

\Received{}
\Accepted{}
 
\usepackage{lineno}

\begin{document} 

\title{On the X-ray efficiency of the white dwarf pulsar candidate ZTF~J190132.9+145808.7
}

\author{Aya \textsc{Bamba}\altaffilmark{1,2,3}}
\author{Yukikatsu \textsc{Terada}\altaffilmark{4}}
\author{Kazumi \textsc{Kashiyama}\altaffilmark{5,6}}
\author{Shota \textsc{Kisaka}\altaffilmark{7}}
\author{Takahiro \textsc{Minami}\altaffilmark{1,6}}
\author{Tadayuki \textsc{Takahashi}\altaffilmark{6,1}}

\altaffiltext{1}{Department of Physics, Graduate School of Science,
The University of Tokyo, 7-3-1 Hongo, Bunkyo-ku, Tokyo 113-0033, Japan}
\email{bamba@phys.s.u-tokyo.ac.jp}
\altaffiltext{2}{Research Center for the Early Universe, School of Science, The University of Tokyo, 7-3-1
Hongo, Bunkyo-ku, Tokyo 113-0033, Japan}
\altaffiltext{3}{Trans-Scale Quantum Science Institute, The University of Tokyo, Tokyo  113-0033, Japan}
\altaffiltext{4}{Graduate School of Science and Engineering, Saitama University, 255 Shimo-Ohkubo, Sakura, Saitama 338-8570, Japan}
\altaffiltext{5}{Astronomical Institute, Tohoku University, Sendai 980-8578, Japan}
\altaffiltext{6}{Kavli Institute for the Physics and Mathematics of the Universe, The University of Tokyo, Kashiwa 277-8583, Japan}
\altaffiltext{7}{Physics Program, Graduate School of Advanced Science and Engineering, Hiroshima University, Higashi-Hiroshima 739-8526, Japan}



\maketitle

\begin{abstract}
Strongly magnetized, rapidly rotating massive white dwarfs (WDs) emerge as potential outcomes of double degenerate mergers.
These WDs can act as sources of non-thermal emission and cosmic rays, gethering attention as WD pulsars.
In this context, we studied the X-ray emissions from ZTF~J190132.9+145808.7 (hereafter ZTF~J1901+14),
a notable massive isolated WD in the Galaxy,
using the Chandra X-ray observatory.
Our results showed 3.5$\sigma$ level evidence of X-ray signals, although it is marginal.
Under the assumption of a photon index of 2, we derived its intrinsic flux to be 2.3 (0.9--4.7) $\times 10^{-15}$~erg~cm$^{-2}$s$^{-1}$ and luminosity  4.6 (2.0--9.5) $\times 10^{26}$~erg~s$^{-1}$ for a 0.5--7~keV band in the 90\% confidence range,
given its distance of 41~pc.
We derived an X-ray efficiency ($\eta$) concerning the spin-down luminosity to be 0.012 (0.0022–-0.074),
a value comparable to that of ordinary neutron star pulsars.
The inferred X-ray luminosity may be compatible with 
curvature radiation from sub-TeV electrons accelerated 
within open magnetic fields in the magnetosphere of ZTF J1901+14.
Conducting more extensive X-ray observations is crucial 
to confirm whether ZTF J1901+14-like isolated WDs are also significant sources of 
X-rays and sub-TeV electron cosmic rays, similar to other WD pulsars in accreting systems.

\end{abstract}


\section{Introduction}

Low-mass stars like our Sun evolve into white dwarfs (WDs) at the end of their lifetimes.
One-third of stellar objects are believed to be white dwarfs.
These objects are significant not only as major constituents of the Galaxy 
but also as key entities for understanding physics under high density environments,
the mechanism of SN Ia explosions, and more.
Additionally, WDs close to the Chandrasekhar limit are pivotal for the substantial production of neutronized species like $^{58}$Ni and $^{55}$Mn through efficient electron capture processes
\citep{iwamoto1999,seitenzahl2013}.

Massive and rapidly rotating WDs are of particular interest
because they can form through WD-WD mergers \citep{dan2014}.
As a result of such mergers, these massive WDs possess a smaller radius and higher density, consistent with their equation of states
($M R^{1/3} \approx \mathrm{const}.$,
where $M$ and $R$ denote mass and radius respectively; \citep{schwab2021}). 
A more rapid rotation period is anticipated due to the conservation of angular momentum,
although the extent of this conservation remains a topic of study \citep{schwab2021}.
Their dipole magnetic fields are believed to be intensified by potent dynamo mechanisms during mergers
\citep{tout2008,garcia2012,das2012}.

Identifying and quantifying such massive and rapidly rotating WDs are crucial steps
in understanding WD-WD merger rates and their associated nucleosynthesis processes.
While massive WDs typically exhibit high-temperature colors in the optical band, most have spin periods around $10^4$~s or longer.
Recent deep optical surveys have identified several potential remnants of WD-WD mergers with rapid spin periods.
One notable candidate identified by the Zwicky Transient Facility is ZTF J190132.9+145808.7 (hereafter ZTF~J1901+14).
This WD is near the Chandrasekhar limit with a mass between 1.327--1.365~$M_\odot$,
a measured radius comparable to the Moon,
and a notably short spin period of 416~s
\citep{caiazzo2021}.

We propose an innovative approach to investigate these magnetic and rapidly rotating WDs.
They are expected to emit nonthermal X-rays as a result of their spin-down, 
akin to isolated neutron stars, leading to their nickname ``white dwarf pulsars".
Such WDs are anticipated to release hard pulsating X-rays, presenting a novel method for their identification.
\citet{orstriker1970} originally proposed this idea,
and \citet{sousa2022} also mentioned this possibility.
Observationally, the first such X-ray emission was reported from an accreting magnetized WD, AE Aqr \citep{terada2008},
which has an incredibly fast rotation period of 33~s.
This was followed by discoveries in AR Sco \citep{buckley2017,takata2018} with a period of 118~s and J191213.72$-$441045.1 \citep{pelisoli2023} with a period of 5.30~min.
Their rapid spin, significant magnetic field, and larger radius compared to neutron stars enable them to achieve electrostatic potentials,
$V \propto P^{-2}BR^{3/2}$, comparable to those of neutron stars.
Consequently, WD pulsars can accelerate particles to as high energies as neutron stars.
The X-ray luminosity from these WDs is around 0.1\% of their spin-down energy \citep{terada2008},
similar in efficiency to neutron stars \citep{kargaltsev2008}.
Given the abundance of WDs compared to neutron stars, they might significantly contribute to the Galactic cosmic ray electron-positron components \citep{kashiyama2011,kamae2018}.
Notably, some radio sources labeled "ultra-long period pulsars" could be white dwarfs, providing further evidence of white dwarf pulsars 
\citep{hurleywalker2022,hurleywalker2023}.

However, all identified WD pulsars to date are part of accreting systems. Earlier hard X-ray observations of isolated WDs failed to detect significant nonthermal emissions \citep{harayama2013},
and the constraints from these observations were somewhat loose.
A focused search for isolated WD pulsars is necessary to validate our proposed scenario.
In this paper, we present the first nonthermal X-ray search for the massive, high-magnetic field WD, ZTF J1901+14, using the high-resolution capabilities of the Chandra observatory.
In Section~\ref{sec:obs}, we detail our target selection, observation methods, and data reduction. Section~\ref{sec:result} outlines our imaging and spectral analysis results, while Section~\ref{sec:discuss} offers a discussion of our findings.

\section{Target Selection and observations}    
\label{sec:obs}

Our primary goal is to detect X-rays emitted from isolated, massive, and magnetic white dwarfs (WDs) that exhibit short spin periods.
Recent optical surveys, such as the Zwicky Transient Facility \citep{bellm2019} and the Sloan Digital Sky Survey \citep{eisenstein2006}, have identified several isolated WDs that rotate rapidly. 
Out of 25 WDs from \citet{kilic2023}, which catalogs ultramassive WDs, 
we singled out four as potential WD pulsar candidates,
based on their strong magnetic fields and rapid rotation.
Table~\ref{tab:candidates} showcases the physical parameters of these candidates, juxtaposed with EUVE~J0317$-$855, which has been previously investigated as a WD pulsar candidate \citep{harayama2013}.

Dipole moments ($\mu$) and spin-down energy ($\dot{E}$) were estimated, assuming the relationships
$\mu \propto BR^3$ and $\dot{E} \propto \mu^2P^{-4} \propto B^2R^6P^{-4}$,
where $R$ represents their radius and $P$ is their rotation period.
We utilized \citet{nauenberg1972} to determine $R$, as detailed in Table~\ref{tab:candidates}.
It is worth noting that strong magnetic fields might induce minor variations in $R$, but for our approximate calculations, such effects are negligible.
Table~\ref{tab:candidates} also presents the derived $\dot{E}$, normalized by the value for EUVE~J0317$-$855.

A crucial parameter for detecting significant X-ray emissions is the spin-down flux,
denoted by $\dot{E}/(4\pi d^2)$,
where $d$ is the distance to the target (refer also to \citet{shibata2016,watanabe2019,bamba2020}).
As observed, ZTF~J1901+14 and SDSS~221141.80+113604.5 are poised to exhibit the highest spin-down flux among the five WD pulsar candidates,
making them ideal subjects for our investigation.
However, only ZTF J1901+14 has undergone observations with X-ray observatories, leading us to choose it as our primary target.

\begin{table}
  \tbl{Properties of WD pulsar candidates.}{%
  \begin{tabular}{lccccc}
      \hline
 & ZTF~J1901+14 & J032900.79$-$212309.24\footnotemark[$*$] & J070753.00+561200.25\footnotemark[$*$] & J221141.80+113604.5\footnotemark[$*$] & 
EUVE~J0317$-$855 \\
      \hline
Distance $d$ (pc) & 41 & 59 & 87 & 69 & 27 \\
Mass ($M_\odot$) & 1.327–1.365 & 1.344 & 1.291 & 1.27 & 1.34$\pm$0.3 \\
Radius $R$ (km) & 2140 & 2366\footnotemark[$\dag$] & 2978\footnotemark[$\dag$] & 3194\footnotemark[$\dag$] & 2417\footnotemark[$\dag$] \\
Period $P$ (s) & 416 & 558 & 3780 & 70 & 725 \\
Magnetic field $B$ (MG) & 600--900 & 50--100 & no data & 15 & 450 \\
Dipole moment $\mu$\footnotemark[$\ddag$] & 0.93--1.39 & 0.10--0.21 & no data & 0.08 & 1 \\
Spin-down energy $\dot{E}$\footnotemark[$\ddag$] & 7.9--17.8 & 0.03--0.12 & no data & 68 & 1 \\
Spin-down flux $\dot{E}/4\pi d^2$ & 3.4--7.7 & 0.006--0.02 & no data & 10.4 & 1 \\
References & (1) & (2) & (2) & (3) & (4)(5) \\
\hline
    \end{tabular}}\label{tab:candidates}
\begin{tabnote}
\footnotemark[$*$] SDSS name.  \\
\footnotemark[$\dag$] Estimated with the best-fit value and \citet{nauenberg1972}. \\
\footnotemark[$\ddag$] Normalized to $\dot{E}$ of EUVE~J0317$-$855.\\
Note --- (1) \citet{caiazzo2021}, (2) \citet{kilic2023}, (3) \citet{kilic2021}, (4) \citet{kawka2007}, (5) \citet{harayama2013}
\end{tabnote}
\end{table}


ZTF~J1901+14 was observed by Chandra ACIS-I \citep{weisskopf2002}
on 2022/12/09--10
(OBSID: 26496, 27596, and 27597).
The data reduction and analysis was done with CIAO 4.15 \citep{fruscione2006} and CALDB version of 4.10.4.
We made the reprocessed cleaned data with the standard method following the CIAO guide,
and the resultant exposure time is 39.3~ks.

\section{Results}
\label{sec:result}

\begin{figure}
 \begin{center}
\includegraphics[width=8cm]{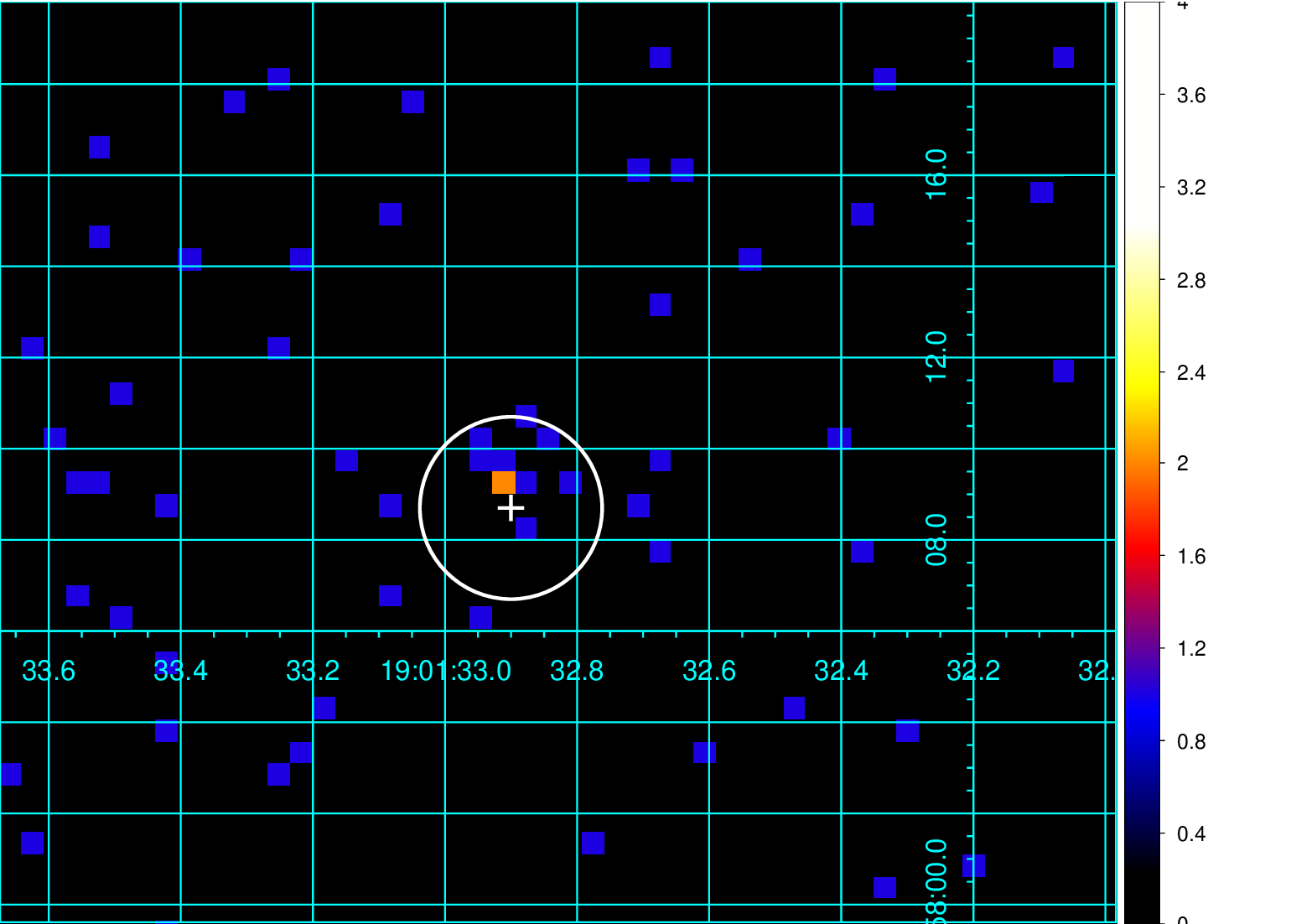}
 \end{center}
\caption{All-band count image of ZTF~J1901+14 region. No vignetting correction has been applied. The scale is in linear, and the coordinates are in J2000. The white cross and circle represent the cataloged position of ZTF~J1901+14 by SIMBAD and its error range with the radius of 2~arcsec.}\label{fig:image}
\end{figure}

Figure~\ref{fig:image} displays the 0.5--7~keV image of the ZTF~J1901+14 region.
In this figure, the position of our target, as determined by SIMBAD \citep{wenger2000}, is marked with a white cross.
Note that the pixel scale of the CCD onboard ZTF is 1~arcsec and the median delivered image quality is 2.0~arcsec in full width at half maximum\footnote{https://www.ztf.caltech.edu/ztf-camera.html}.

There is some excess emission around the target position.
However, it is not significant, and the {\tt wavdetect} command in CIAO did not detect any source in this region.
In order to estimate the flux of the possible X-ray emission on this position,
we cited the coordinate of the pixel which has the maximum count, (RA, Dec.) = (19:01:32.91, +14:58:09.1).
Note that the position is well close to the SIMBAD position
as shown in Figure~\ref{fig:image}.
The significance level of this position is estimated
with {\tt srcflux} command in CIAO
to be 0.998 or 3.5$\sigma$ level.
To convert the count rate to flux, a spectral model is required. We opted for the absorbed power-law model,
which is commonly used for both WD pulsars and neutron star pulsars.
The absorption column was fixed at  $1\times 10^{20}$~cm$^{-2}$,
based on the presumption that the interstellar medium density is 
1~cm$^{-3}$ and considering a distance of 41~pc \citep{gaia2020}.
The photon index of WD pulsars ranges from 1--2.5~\citep{terada2008,takata2018,schwope2023}.
Based on this, we adopted a value of 2. This value is also typical for neutron star pulsars \citep{kargaltsev2008}.
The resulting unabsorbed flux is
2.4 (0.9--4.8)$\times 10^{-15}$~erg~s~cm$^{-2}$ in the 0.5--7~keV band,
where the error range is in 90\% confidence level.
Adjusting the photon index to 1.5 did not lead to significant changes (only less than a few 10\%) in our findings.

\section{Discussion}
\label{sec:discuss}

In the previous section, we presented that the strongly magnetized and rapidly rotating WD, ZTF~J1901+14, has been detected, although it is not prominently bright in the X-ray band.
Assuming a distance of 41~pc, the 90\% error range of the 0.5--10~keV luminosity is 5.3 (2.3--10.9) $\times 10^{26}$~erg~s$^{-1}$.
It is three orders of magnitude lower than the nonthermal X-rays detected from other accreting white dwarf pulsars,
which have X-ray luminosities of approximately $10^{29}$~erg~s$^{-1}$
\citep{terada2008,takata2018,schwope2023},
as listed in Table~\ref{tab:comparison}.

\begin{table}
  \tbl{Properties of WD pulsars and ZTF~J1901+14.}{%
  \begin{tabular}{p{7pc}ccccc}
      \hline
 & AE Aqr & AR~Sco & J191213.72$-$441045.1 & EUVE~J0317$-$855 & ZTF~J1901+14 \\
      \hline
Type \dotfill & accreting & accreting & accreting & isolated & isolated \\
Distance (pc) \dotfill & 92 & 117 & 237 & 27 & 41 \\
$P$ (s) \dotfill & 33 & 118 & 319 & 725 & 416 \\
$B$ (MG) \dotfill & 50 & 900 & no data & 450 & 600--900 \\
$\dot{E}$ (erg~s$^{-1}$) \dotfill & $6\times 10^{33}$\footnotemark[$*$] & $5\times 10^{33}$ & no data & $1.3\times 10^{27}$\footnotemark[$\sharp$]  & (2.9--6.6)$\times 10^{28}$\footnotemark[$\dag$] \\
$\Gamma$ \dotfill & 1.12 & 2.3 & 2.14 & 2.5 (assumed) & 2 (assumed) \\
$F_X$\footnotemark[$\ddag$] (erg~s$^{-1}$cm$^{-2}$) \dotfill & $5.9\times 10^{-13}$ & $2.8\times 10^{-13}$ & $1.3\times 10^{-13}$ & $<4.9\times 10^{-13}$\footnotemark[$\S$] & 2.6 (1.0--5.3) $\times 10^{-15}$\footnotemark[$**$] \\
$L_X$ (erg~s$^{-1}$)\footnotemark[$\|$] \dotfill & $6.0\times 10^{29}$ & $4.6\times 10^{29}$ & $9.1\times 10^{29}$ & $<4.3\times 10^{28}$\footnotemark[$\S$] & 5.3 (2.3-10.9) $\times 10^{26}$\footnotemark[$**$] \\
$\eta \equiv L_X/\dot{E}$ \dotfill & $1.0\times 10^{-4}$ & $9\times 10^{-5}$ & no data & $<$33 & 0.012 (0.0022--0.074)\footnotemark[$\dag\dag$] \\
References & (1)(2) & (3)(4)(5) & (3)(6) & (7)(8) this work & (3)(9) this work \\
\hline
    \end{tabular}}\label{tab:comparison}
\begin{tabnote}
\footnotemark[$*$] Adopted from \citet{dejager1994} and \citet{dejager1994b}.  \\ 
\footnotemark[$\sharp$] With the assumption of the radius of 2417~km (see text). \\
\footnotemark[$\dag$] 
Estimated with \citet{suto2023} (see text).
\\
\footnotemark[$\ddag$] Unabsorbed flux in the 0.5--10~keV band. \\ 
\footnotemark[$\S$]  3$\sigma$ upper limit. \\ 
\footnotemark[$\|$]  In the 0.5--10~keV band. \\ 
\footnotemark[$**$]  90\% error range. \\ 
\footnotemark[$\dag\dag$]  90\% error range assuming the uncertainty of $\dot{E}$ is also in 90\% error range.\\ 
Note --- (1) \citet{terada2008}, (2) \citet{steinmetz2020}, (3) \citet{gaia2020}, (4) \citet{takata2018}, (5) \citet{pelisoli2022}, (6) \citet{schwope2023}, (7) \citet{kawka2007}, (8) \citet{harayama2013} (9) \citet{caiazzo2021}
\end{tabnote}
\end{table}

\begin{figure}
 \begin{center}
\includegraphics[width=0.4\textwidth]{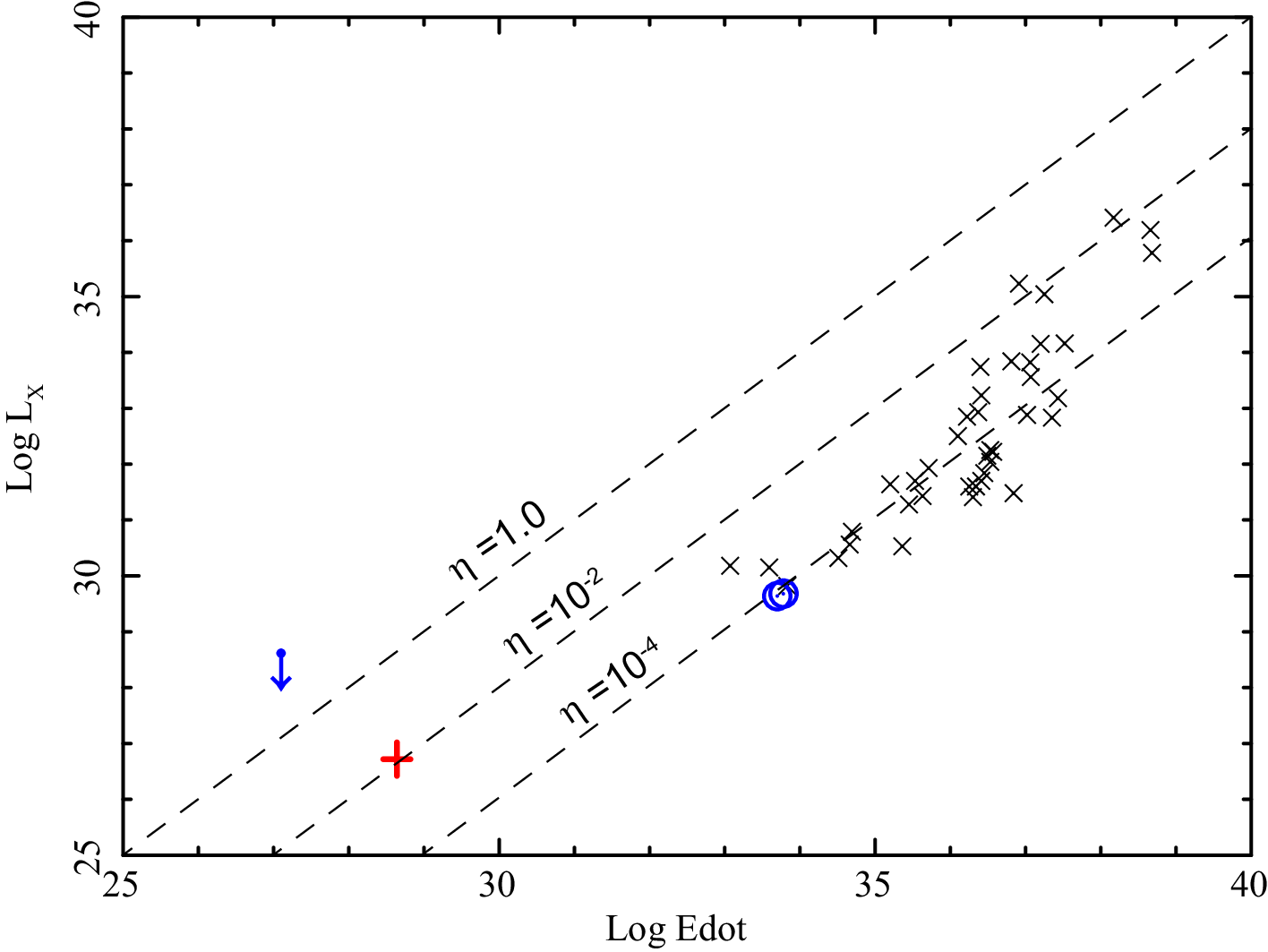} 
 \end{center}
\caption{Spin down energy vs. 0.5--8~keV X-ray luminosity of rotation powered neutron stars (black X; \cite{kargaltsev2008}), accreting WD pulsars AE~Aqr and AR~Sco (blue circles; \cite{terada2008,takata2018}), an isolated WD EUVE~J0317$-$855 (blue upper-limit; \cite{harayama2013}), and ZTF~J1901+14 (in red).
The upper limits are all in 3$\sigma$ level,
and the error range for ZTF~J1901+14 is in 90\%.
Note that the uncertainty is large with just 3.5$\sigma$ detection level.
The spin-down energy for EUVE~J0317$-$855 is
estimated with the WD radius of 2417~km (see table~\ref{tab:candidates}.
Dashed lines represent
$\eta$ of 1, $1\times 10^{-2}$, and $1\times 10^{-4}$ from top to bottom, respevtively.
}\label{fig:Edot-LX}
\end{figure}

In the case of neutron stars, the spin-down energy $\dot{E}$ and X-ray efficiency $\eta$, which is defined as (X-ray luminosity)/$\dot{E}$,
serve as crucial parameters for their understanding
(e.g., \cite{kargaltsev2008}).
We thus evaluated $\dot{E}$ and $\eta$ for our samples and made comparisons.
\citet{caiazzo2021} measured the surface magnetic field $B_s$ and the radius $R$ of ZTF~J1901+14 to be between 600--900~MG and $2140^{+160}_{-230}$~km, respectively.
Meanwhile, \citet{suto2023} proposed that the magnetic field structure of ZTF~J1901+14 is dipolar,
with the magnetic axis inclined at $\chi = $60~degrees to the rotation axis, based on optical/UV light curve analyses.
Assuming the magnetosphere is force free,
$\dot{E}$ is estimated to be
\begin{equation}
\dot{E} = \mu^2\left(\frac{2\pi}{P}\right)^4c^{-3}(1+C\sin^2(\chi))
\sim (1.5-10) \times 10^{28}~{\rm erg~s^{-1}},
\end{equation}
where $\mu = B_{\rm d}R^{3}/2$ with the dipole magnetic field $B_{\rm d}$, with the assumption of $B_{\rm d} = B_{\rm s}$, $c$ is the light speed, and the constant $C \sim 1$
\citep{gruzinov2005,spitkovsky2006,tchekhovskoy2013}.
For accreting WD pulsars, the $\dot{E}$ of AE~Aqr was taken to be
$6\times 10^{33}$~erg~s$^{-1}$, as estimated by \citet{dejager1994} and \citet{dejager1994b}.
For AR~Sco, we utilized the spin-down frequency of
$\dot{\nu} = 4.47\times 10^{-17}$~Hz~s$^{-1}$ \citep{pelisoli2022}, resulting in $\dot{E} = 5\times 10^{33}$~erg~s$^{-1}$ and $B_d = 900$~MG, with assumptions regarding that the spin-down is due to magnetic dipole radiation, $M = 0.8~M_\odot$, and $R =$ 7000~km. 
The derived parameters are also listed in Table~\ref{tab:comparison}.
Regarding X-ray efficiency, $\eta$,
AE~Aqr and AR~Sco are approximately $1\times 10^{-4}$,
whereas ZTF~J1901+14 is 
0.012 ($0.0022-0.074$)
In contrast, typical neutron stars have $\eta$ values between $10^{-5}$ and $10^{-1}$ \citep{kargaltsev2008}.
We can thus infer that its X-ray efficiency 
is similar to typical neutron stars,
although the uncertainty is rather large.
Figure~\ref{fig:Edot-LX} plots the relationship between $\dot{E}$ and X-ray luminosity for our samples compared to neutron stars, further illustrating this point.

We also drew comparisons with another isolated WD pulsar candidate,
EUVE~J0317$-$855.
\citet{harayama2013} estimated its $\log\dot{E}$ 29.0--30.8
with the assumption of the radius of 5000--10000~km.
On the other hand, the mass of EUVE~J0317$-$855 is 1.31--1.37~M$_\odot$
\citep{kawka2007},
which is quite similar to that of ZTF~J1901+14,
and its radius should be as small as that of ZTF~J1901+14
according to the mass-radius relation of WDs.
We thus re-estimated its $\log\dot{E}$
to be 27.1 (or $\dot{E} = 1.3\times 10^{27}$~erg~s$^{-1}$)
with the assumed radius of 2417~km (see table~\ref{tab:candidates}),
following that the magnetic dipole moment $\mu \propto BR^3$ and $\dot{E} \propto \mu^2P^{-4} \propto B^2R^6P^{-4}$.
The re-estimated parameters for EUVE~J0317$-$855 are also listed in Table~\ref{tab:comparison}.
With this $\dot{E}$, the 3$\sigma$ upper-limit of the X-ray efficiency for EUVE~J0317$-$855 is 33
in the 0.5--8~keV band,
which is much larger the value with larger radius assumptions.
Figure~\ref{fig:Edot-LX} also shows the EUVE~J0317$-$855,
with the assumptions of our new radius estimation.
The figure shows clearly that
the X-ray efficiency of ZTF~J1901+14 is well constrained
compared with that for EUVE~J0317$-$855,
although the significance is not so high.

Here, we note the uncertainty of our results.
One of the concern is the uncertainty of the derived radius with the model by \citet{nauenberg1972},
which ignored the effect of magnetic field and rotation:
the strong magnetic field makes degenerate pressure smaller and the stellar radius becomes larger,
and as a result, the derived spin-down energy becomes larger
and the X-ray efficiency smaller.
This effect clearly appear the magnetic field much larger than$\sim 10^{9}$ Gauss,
thus in this sense, our conclusion (not very bright in X-rays) does not change.
The error range of the radius for ZTF~J1901+14 is just 10\% \citep{caiazzo2021},
which is already included in our discussion.
The distance uncertainty is negligible, less than 0.1~pc
\citep{gaia2020}.

Assuming that the X-ray detection from ZTF~J1901+14 is the case, 
let us explore the potential emission mechanism. 
Given the measured radius, magnetic field strength, and rotation period, 
ZTF~J1901+14 is likely below the death line for WD pulsars~\citep{kashiyama2011}, 
i.e., electron-positron pair multiplication process in the magnetosphere 
will not be operational.
Nevertheless, charged particle could still undergo acceleration 
through the electric field induced by the unipolar induction, 
especially along the open field lines. 
The maximum voltage can be estimated as 
\begin{equation}
    V_{\rm max} \approx \frac{B_\mathrm{d}(2\pi/P)^2R^{3}}{2c^{2}} \sim 4.2\times 10^{11}\,{\rm Volt}.
\end{equation}
Hereafter, we adopt the maximum values of the radius and magnetic field strength within uncertainties~\citep{caiazzo2021} as our fiducial values.
Accordingly, electrons could be accelerated to sub-TeV energies 
with Lorentz factors of $\gamma_{\rm max} \approx eV_\mathrm{max}/m_\mathrm{e}c^2 \sim 8.1\times 10^5$.
Assuming that electrons with a Goldreich-Julian density $n_\mathrm{GJ} \approx B_\mathrm{d}/ce P$ are supplied quasi-steadily into the polar cap region $r_\mathrm{cap} \approx R \times (2\pi R/cP)^{1/2}$,
the kinetic luminosity of the electrons becomes comparable to the spindown luminosity, i.e., $L_\mathrm{e} \approx 2\pi r_\mathrm{cap}^2 c m_\mathrm{e} n_\mathrm{GJ} \gamma_\mathrm{max} c^2 = 2 \mu^2 (2\pi/P)^4c^{-3} \approx \dot E$.
These electrons partially lose their energies through the curvature radiation. 
The emission frequencies can be estimated as 
\begin{equation}
h\nu_\mathrm{c} \approx \frac{3h\gamma_{\rm max}^3c}{4\pi R_\mathrm{c}} \sim 69\,\mathrm{keV}\,\left(\frac{R_\mathrm{c}}{R}\right)^{-1}
\end{equation}
where $h$ is the Planck constant, 
$R_\mathrm{c}$ is the curvature radius of the open field lines.
In the case of ZTF~1901+14, the emission is predominantly in the X-ray band 
for $R \lesssim R_\mathrm{c} \lesssim 100 R$, i.e., in the near surface region.
The X-ray luminosity can be estimated as $L_X \approx (t_\mathrm{c}/t_\mathrm{dyn}) \times L_\mathrm{e} \approx (t_\mathrm{c}/t_\mathrm{dyn}) \times \dot E$, where $t_\mathrm{dyn} \approx R/c$ is the dynamical timescale of the electrons and $t_\mathrm{c} \approx 3m_\mathrm{e}cR_\mathrm{c}^2/2e^2\gamma_\mathrm{max}^3$ is the energy loss timescale through the curvature radiation.  
Then, the X-ray radiation efficiency is given as
\begin{equation}
\eta = \frac{L_X}{\dot E} \approx \frac{t_\mathrm{dyn}}{t_\mathrm{c}} \sim 1.3\times 10^{-3} \left(\frac{R_\mathrm{c}}{R}\right)^{-2}.
\end{equation}
Hence, within the uncertainties, the observed X-ray luminosity of ZTF~J1901+14 may be attributed to the curvature radiation from sub-TeV electrons accelerated along the open field lines.

Since $\eta \ll 1$, sub-TeV electrons escape into the interstellar medium without significant energy loss, thus supporting the idea that strongly magnetized, rapidly rotating massive WDs like ZTF J1901+14 are efficient factories for electron cosmic rays~\citep{kashiyama2011}. 
The lower X-ray efficiency compared to X-ray-bright neutron star pulsars can be attributed to its relatively low-density environment. 
Relativistic winds and cosmic rays launched from the magnetosphere of ZTF J1901+14 would be directly injected into the interstellar medium, whereas those from a young neutron star pulsar are surrounded by a supernova remnant~(e.g., \cite{gelfand2009,bamba2010}) and can efficiently dissipate at and beyond the wind termination shock~(e.g., \cite{meintjes2023}). 
Accreting WD pulsars are typically found in environments with a relatively high density of materials from their companion stars, as observed in the case of AE Aqr~\citep{terada2008}, or they may encounter stellar winds from their companion stars, as seen in the case of AR Sco~\citep{buckley2017,takata2018}.

Further investigations, such as deeper X-ray observations of ZTF J1901+14 and measurements of the magnetic field of J191213.72$-$441045.1, will provide more insights into clarifying the differences in non-thermal emissivity among these types of compact star systems.
We need deeper X-ray observations
with large effective area missions
such as Athena \citep{nandra2013} and Lynx \citep{gaskin2018},
in order to make quantitative comparison of X-ray efficiencies of other compact star systems.
The requirement of point source sensitivity of Wide Field Imager (WFI) onboard Athena
is in the order of $10^{-17}$~erg~s$^{-1}$cm$^{-2}$ \citep{nandra2013},
resulting $\eta \sim 10^{-4}$ for ZTF~J1901+14, 
which is enough to judge isolated massive and rapid-rotating WDs are
efficient X-ray emitters or not.

\section{Summary}\label{sec:summary}

We conducted an X-ray search for ZTF~J1901+14,
one of the most massive and rapidly-rotating white dwarfs, using Chandra
and found X-ray emission with the significence of 3.5$\sigma$ level.
Assuming a photon index of 2, 
we derived its intrinsic flux to be 2.3 (0.9--4.7) $\times 10^{-15}$~erg~cm$^{-2}$s$^{-1}$ and luminosity  4.6 (2.0--9.5) $\times 10^{26}$~erg~s$^{-1}$ for a 0.5--7~keV band in the 90\% confidence range,
given its distance of 41~pc.
We derived an X-ray efficiency ($\eta$) concerning the spin-down luminosity to be 0.012 (0.0022–-0.074),
which is
similar to typical neutron star pulsars.
The observed X-ray emission might suggest that such WDs have the capability to accelerate electrons to sub-TeV energies.
However, conclusive evidence requires more in-depth observations.
For a comprehensive comparison with other WD pulsars that have companion stars, like AE~Aqr, AR~Sco, and SDSS~J191213.72$-$441045.1,
further X-ray/optical observations are essential to pinpoint physical parameters including the magnetic field and spin-down luminosity.

\begin{ack}
We thank the anonymous referee for his/her constructive comments.
We thank Shinpei Shibata for the fruitful discussions.
This research has made use of the SIMBAD database,
operated at CDS, Strasbourg, France.
This work was financially supported by Japan Society for the Promotion of Science Grants-in-Aid for Scientific Research (KAKENHI) Grant Numbers, JP19K03908 (AB), JP23H01211 (AB), JP20K04009 (YT), JP20H01904 (KK), JP22H00130(KK), JP23H04899 (KK), JP21H01078 (SK), JP22H01267 (SK), JP22K03681 (SK), and JP20H00153 (TT).
\end{ack}


\end{document}